\documentclass[prl,aps,twocolumn]{revtex4}
\usepackage{graphicx}
\usepackage{latexsym}
\usepackage{amsmath}
\begin{document}

\title{Infinite randomness fixed point of the 
superconductor-metal quantum phase transition}

\author{Adrian Del Maestro}
\affiliation{Department of Physics, Harvard University, Cambridge,
MA 02138}

\author{Bernd Rosenow}
\affiliation{Department of Physics, Harvard University, Cambridge,
MA 02138}

\author{Markus M\"uller}
\affiliation{Department of Physics, Harvard University, Cambridge,
MA 02138}

\author{Subir Sachdev}
\affiliation{Department of Physics, Harvard University, Cambridge,
MA 02138}

\date{\today}

\begin{abstract}
We examine the influence of quenched disorder on the superconductor-metal transition, as described
by a theory of overdamped Cooper pairs which repel each other.
The self-consistent pairing eigenmodes of a quasi-one dimensional wire are determined numerically.
Our results support the recent proposal by Hoyos {\em et al.\/} (Phys. Rev. Lett. {\bf 99}, 230601
(2007)) that the transition is characterized by the same strong disorder
fixed point describing the onset of ferromagnetism in the random quantum Ising chain in a 
transverse field.
\end{abstract}
\maketitle

Numerous recent experiments
\cite{liu-zadorozhny,lau-markovic,boogaard,rogachev-bollinger,chang,rogachev-wei}
have measured the electrical transport properties
of quasi-one dimensional nanowires. While thicker wires have vanishing resistance in the
low temperature ($T$) limit, thinner wires do not display superconductivity even at the lowest
$T$. The superconducting wires display clear signatures of thermal phase fluctuations of
the Cooper pair order parameter, $\Psi$, at low $T$. Quantum fluctuations of the
phase and amplitude of $\Psi$ increase with decreasing wire thickness, leading to a
transition to a non-superconducting state.

Recent work \cite{swt,mrss} has proposed that these experiments should be
described by a quantum superconductor-metal transition (SMT) in the pair-breaking universality
class. Arguments based upon microscopic
BCS theory were used to propose a model of $\Psi$ fluctuations damped by decay into single electron
excitations of the metal \cite{herbut,lsv,spivak-zyuzin,fl,galitski}.
In this paper, we will present the results of a numerical study of the influence
of quenched disorder on this model. The role played by disorder near the quantum SMT is of
considerable interest, as disorder correlations are of infinite range in the imaginary time
direction and can lead to unusual critical phenomena \cite{vojta-review}.

In a renormalization group (RG) analysis of this
overdamped Cooper pair model with disorder, Hoyos {\em et al.} \cite{hoyos} have recently argued
that the SMT is described by a strong disorder fixed point.  The latter exhibits activated dynamic 
scaling where the logarithm of characteristic frequencies of $\Psi$ fluctuations grows as a
power of their characteristic length scale. They argued further that the strong disorder fixed 
point is in the same universality class as the one describing the onset of ferromagnetism in the 
quantum random transverse field Ising model (RTFIM) in one spatial dimension. Many 
exact results were obtained by Fisher \cite{fisher} for this fixed point, and carry through to a 
large extent to the case of the SMT. Note that this is a non-trivial result, since the RTFIM contains {\em no dissipation\/}, 
and possesses a {\em discrete symmetry}. All our numerical results below confirm the remarkable 
predictions of Hoyos {\em et al.\/}, providing strong evidence for the applicability of their 
strong randomness RG. 

Our analysis was carried out on a lattice discretization of the disordered overdamped
Cooper pair model of Ref.~\onlinecite{hoyos} at $T = 0$. The degrees of freedom are $\Psi_j 
(\tau)$, which are complex functions of imaginary time, $\tau$, on the sites, $j$, of a 
one-dimensional chain with action
\begin{align}
\mathcal{S} &= \sum_j \int d \tau \left[ D_j |\Psi_j - \Psi_{j+1} |^2 + \alpha_j |\Psi_j|^2 +
\frac{u_j}{2} |\Psi_j|^4 \right] \nonumber \\
&\quad+ \int \frac{d\omega}{2\pi} \sum_j \gamma_j |\omega| |\Psi_j (\omega) |^2,
\label{S}
\end{align}
where $\Psi_j (\omega)$ is the Fourier transform of $\Psi_j (\tau)$, and the couplings in 
$\mathcal{S}$ are all random functions of $j$. The quartic coefficients $u_j$ are all positive to 
ensure stability and repulsion between Cooper pairs. The dissipation into the metallic bath is 
represented by $\gamma_j$, which is also required to be positive by causality. Finally, we can 
choose a gauge such that $D_j > 0$. A more careful analysis and
suitable rescalings \cite{tucker} allow us to reduce the randomness to the spatial dependence of $D_j$
(uniformly distributed on $(0,1]$) and $\alpha_j$ (taken to be Gaussian),
while setting $u_j = u$ and $\gamma_j = 1$.  At zero temperature, the SMT can be tuned by reducing 
the mean of the $\alpha_j$ distribution, $\overline{\alpha}$, while keeping its variance constant 
at $0.25$ in units of $\gamma^2$.

Equivalently, we can also work in a lattice model of fluctuating phases
with $\Psi_j (\tau) = e^{i \theta_j (\tau)}$ of unit magnitude \cite{swt,fl}; this
should have the same properties as $\mathcal{S}$, but our analysis proceeds more
conveniently by also allowing for magnitude fluctuations.

While $\mathcal{S}$ is a suitable model for describing the influence of disorder
on the fluctuating Cooper pair states, we also have to consider the effect of randomness on the
single electron states.  We have estimated such effects in the framework of weak-coupling BCS 
theory: at criticality, we find that on a  scale parametrically smaller than the single electron 
localization length, the gain in condensation energy can offset the cost in elastic energy when  
order parameter fluctuations take advantage  of  randomness in the $\alpha_j$. This justifies 
our focus on the influence of disorder in a purely bosonic overdamped Cooper pair theory. 
Details will be provided later (see also Ref.~\onlinecite{spivak-zyuzin}).

The RG analysis \cite{hoyos} was carried out in a model with an $N$-component order parameter
and it was found that flows had only an irrelevant dependence on the value of
$N$ \cite{satya}.  Thus the exact critical properties can be obtained by studying
the model in the large $N$ limit. This is equivalent to approximating $\mathcal{S}$
by the Gaussian action
\begin{equation}
\mathcal{S}_0 = \sum_j\! \int \frac{d\omega}{2\pi} \left[ D_j |\Psi_j - \Psi_{j+1} |^2 + (r_j +
|\omega|) |\Psi_j|^2 \right],
\label{eq:S0}
\end{equation}
where the $r_j$ are determined self-consistently by solving
\begin{equation}
r_j = \alpha_j + u\left\langle |\Psi_j (\tau) |^2 \right\rangle_{\mathcal{S}_0}.
\label{eq:rj}
\end{equation}
We set $u = 1$ to reach a strong coupling regime and use fixed but random boundary conditions, 
similar to those employed in Ref.~\onlinecite{swt}.
Solving the innocuous looking Eq.~(\ref{eq:rj}) for a large number of disorder realizations and 
large system sizes was the primary time-consuming numerical step in obtaining the results of this study.
Similar numerical large-$N$ methods have been used previously for disordered systems
with conventional (power law) dynamic scaling~\cite{tu-weichman,hartman-weichman} but the presence 
of activated scaling leads to sluggish dynamics and the necessity to properly include 
spurious disorder configurations that, although exponentially rare, can make large contributions to 
thermodynamic properties.  The numerical solution is facilitated via the implementation of a method 
which we have dubbed the \emph{solve-join-patch} (SJP) procedure.  We begin by generating a realization 
of disorder for $L$ sites with $L$ large.  Near the critical point, characterized by the condition that 
the correlation length $\xi \sim L$, the direct iterative solution of Eq.~(\ref{eq:rj}) is 
computationally quite costly.  This is a result of the fact that the eigenmodes of ${\mathcal S}_0$ begin 
to delocalize and have a characteristic energy scale that is exponentially
small in the distance from criticality, requiring that the solutions $r_j$ must be computed with
exponentially increasing precision.

To cope with this difficulty, the system consisting of $L$ sites is broken up into
a group of smaller sub-systems with boundary conditions adjusted to reflect their location in the
larger chain.  The sub-systems are \emph{solved}, then \emph{joined} together in groups of two.
The grouped sub-systems are now close to satisfying Eq.~(\ref{eq:rj}) and they can be quickly 
brought into accordance by \emph{patching}, which involves re-solving a mini-system around the joint
consisting of a small number of sites.  The joined and patched sub-system can now be easily solved and
the SJP procedure is iterated until a full solution to  Eq.~(\ref{eq:rj}) is obtained for the
complete chain of $L$ sites.  We have considered up to $3000$ realizations of disorder for system
sizes $L=16, 32, 64$, and $128$.

Fisher's remarkable solution of the RTFIM \cite{fisher} includes asymptotically exact results for 
the exponents and correlation functions at the infinite randomness fixed point, and many 
directly translate to the RG calculations by Hoyos {\em et al.} \cite{hoyos} for the dissipative model
considered here. In particular, one expects activated dynamic scaling with $\ln (1/\Omega) \sim 
L^{\psi}$ where $\Omega$ is a characteristic energy scale and $\psi = 1/2$ is a tunneling exponent. 
This reflects the fact that at an infinite randomness fixed point, the dynamical critical exponent 
$z$ is formally infinite.  The RG approach defines a real space decimation procedure that either 
creates or destroys \emph{clusters} or \emph{bonds} as the energy scale is reduced. 
The typical moment of a surviving cluster scales like 
$\mu \sim \ln^\phi (1/\Omega)$ at criticality, where $\phi = (1+\sqrt{5})/2 \simeq 1.62$ is the 
golden mean.  Average correlations are described by a correlation length which diverges as
$\xi \sim |\delta|^{-\nu}$ with $\nu = 2$ and $\delta$ a measure of the distance from criticality. 
From Ref.~\onlinecite{hoyos}, $\delta$ is expected to be proportional to $\overline{\ln(r_i/r_c)}$ 
where $r_c$ is some critical value. Our numerical study reveals that close to criticality this 
quantity is linearly related to the detuning of the average $\overline{\alpha}$ from its quantum 
critical value, $\overline{\alpha}_c$ (which has yet to be determined) and it further 
demonstrates that correlations among the $r_i$ due to their self-consistency does not affect 
the strong randomness RG flow.

The remainder of this paper will present a numerical confirmation of the results of 
Ref.~\onlinecite{hoyos} by providing arguments for the presence of dynamically activated scaling at 
the quantum SMT, characterized by exponents $\nu$, $\psi$ and $\phi$ taking on their  
RTFIM values. The evidence comes from an analysis of equal time correlations, energy gap 
statistics and dynamic susceptibilities in the weakly disordered quantum Griffiths phase 
\cite{griffiths}.

\begin{figure}[ht]
\centering
\includegraphics*[width=3.3in]{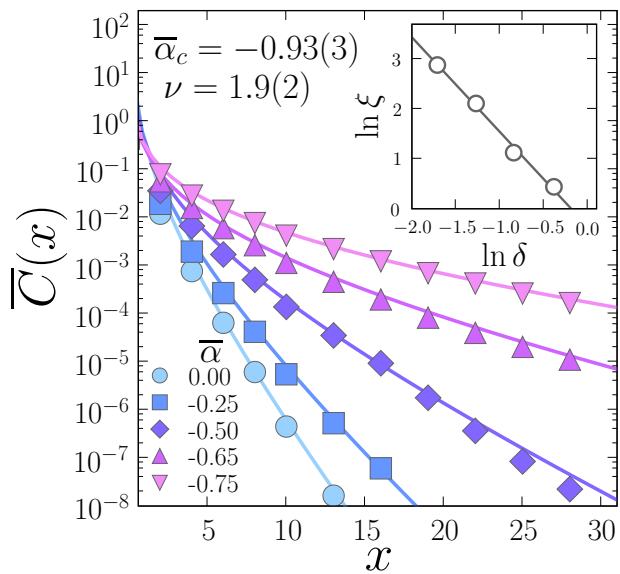}
\caption{\label{fig:eqTimeCorFunc} (Color online) The equal-time disorder averaged correlation 
functions for $L = 64$ and five values of the mean of the $\alpha_j$ distribution, $\overline{\alpha}$.
The solid lines are fits to the asymptotic form described in Eq.~(\ref{eq:fisherEqTimeCorFunc}) via 
$\xi$ and an overall scale parameter.  The inset shows the result of a fit to the power law form of the 
finite size scaled correlation length leading to an estimate for the location of the critical 
point $\overline{\alpha}_c = -0.93(3)$ and the correlation length exponent $\nu = 1.9(2)$.}
\end{figure}

We begin by studying the disorder averaged equal-time correlation function
$\overline{C}(x) = \overline{\langle \Psi_x^*(\tau)\Psi_{0}(\tau)\rangle}_{\mathcal{S}_0}$, which
can be computed from the quadratic effective action $\mathcal{S}_0$ once the full set of solutions
$\{r_j\}$ has been obtained.  In the disordered phase, where $\delta \equiv
\overline{\alpha}-\overline{\alpha}_c > 0$ the asymptotic form of $\overline{C}(x)$ for
the RTFIM has been predicted to describe both exponential as well as stretched exponential decay
in addition to power law behavior \cite{fisher}
\begin{equation}
\overline{C}(x) \sim \frac{\exp\left[-(x/\xi) - (27\pi^2/4)^{1/3}(x/\xi)^{1/3}\right]}
{(x/\xi)^{5/6}}.
\label{eq:fisherEqTimeCorFunc}
\end{equation}
If we use this expression to define the correlation length $\xi$, we can perform fits for each
value of $L$ and various $\overline{\alpha}$ to extract $\xi(L,\overline{\alpha})$ as is seen in
Fig.~\ref{fig:eqTimeCorFunc} for $L = 64$.
We find remarkable agreement (solid lines) with Eq.~(\ref{eq:fisherEqTimeCorFunc}) over six orders of
magnitude for all system sizes considered.

As mentioned above, the length scale which describes average correlations is expected to diverge
like $\xi\sim|\delta|^{-\nu}$ as the critical point is approached. We have employed this result to
perform a log-log fit to the finite size scaled correlation length (data extrapolated to $L\to 
\infty$) as a function of $\delta$, as is shown in the inset of Fig.~\ref{fig:eqTimeCorFunc}.  
The value of $\overline{\alpha}_c$ was found from the mean of the critical $\alpha_j$ distribution
which minimized the least square error of power law fits involving $\delta =
\overline{\alpha}-\overline{\alpha}_c$.
This leads to a value of $\overline{\alpha}_c = -0.93(3)$ for the critical point and $\nu=1.9(2)$ 
for the correlation length exponent with the number in brackets indicating the uncertainty in the 
last digit computed from the fitting procedure.  The obtained value of $\nu$ is in accord with the 
value of $2$ predicted for the RTFIM.  The correlation length could also have been defined via the 
exponential tail of $\overline{C}(x)$ at large separations which yields compatible values for both 
$\overline{\alpha}_c$ and $\nu$.  

\begin{figure}[ht]
\centering
\includegraphics*[width=3.3in]{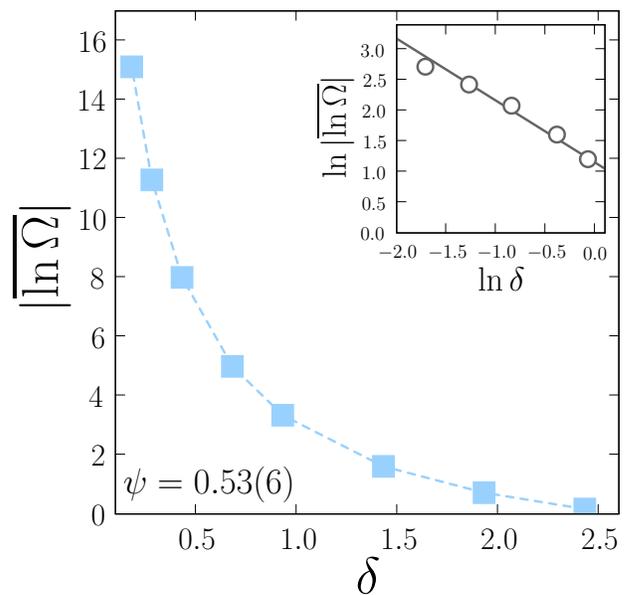}
\caption{\label{fig:lnGap}(Color online) The finite size scaled value of the disorder averaged 
logarithm of the minimum excitation energy plotted against the distance from the critical point 
$\delta$.  We observe divergence consistent with the scaling form 
$|\overline{\ln\Omega}| \sim \delta^{-\nu\psi}$ and using the value of $\overline{\alpha}_c$ and 
$\nu$ found above we determine $\psi = 0.53(6)$ from a log-log linear fit (inset).}
\end{figure}

For each realization of disorder and each value of $\overline{\alpha}$ we define the gap 
$\Omega(L)$ to be the smallest excitation energy in the system, which in general corresponds to the most 
delocalized mode of $\mathcal{S}_0$.  Rare disorder configurations cause clusters to behave as if 
they were much more critical than the global value of $\delta$ would suggest.  These clusters dominate 
the critical modes and exhibit abnormally small gaps that make large contributions to disorder 
averages of $\ln \Omega$, leading to the highly anisotropic scaling relationship between space and time 
that is the hallmark of strong disorder fixed points. 
An analysis of the probability distribution for the logarithm of the energy gap in the RTFIM was carried 
out by Young and Rieger \cite{young-rieger} where they found cogent evidence for $z = \infty$.
We have performed a similar analysis here, with the same result. In addition we find that 
$\ln \Omega$, as the minimal excitation energy, is naturally Gumbel distributed.
If activated dynamic scaling is indeed present, 
the disorder averaged value of the logarithm of the gap should scale like 
$|\overline{\ln\Omega}| \sim \xi^{\psi} \sim \delta^{-\nu\psi}$ where we have used the scaling form 
of the correlation length.  Such divergent behavior for the finite size scaled value
of $|\overline{\ln \Omega}|$ is demonstrated in Fig.~\ref{fig:lnGap}.
The possibility of conventional scaling was considered but ultimately excluded through the 
examination of the maximum likelihood estimator for a wide range of power law fits.  
Using the previously determined values of $\overline{\alpha}_c$ and $\nu$, the tunneling exponent can be
extracted from a log-log linear fit of the average logarithmic spectrum as shown in the inset of
Fig.~\ref{fig:lnGap}, producing $\psi = 0.53(6)$ which is consistent with the RTFIM prediction
of $1/2$.

\begin{figure}[ht]
\centering
\includegraphics*[width=3.3in]{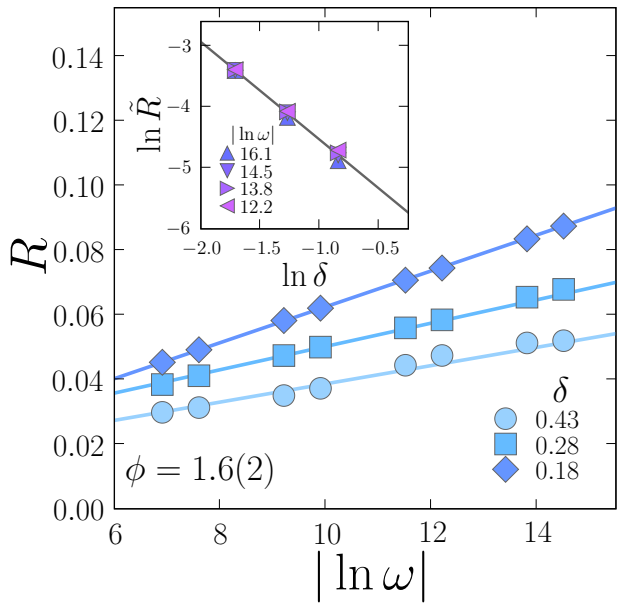}
\caption{\label{fig:R}(Color online) The real frequency dependence of the finite size scaled value 
of the disorder averaged susceptibility ratio defined in Eq.~(\ref{eq:R}) for three values of
the $\delta = \overline{\alpha}-\overline{\alpha}_c$.  We observe the predicted $|\ln\omega|$ 
behavior. After a suitable rescaling described in the text we find that $\widetilde{R}$ does not depend 
on frequency as $\omega \to 0$ (inset), and a log-log linear fit gives the value of the cluster exponent 
to be $\phi = 1.6(2)$.}
\end{figure}

To confirm full agreement with the universality class of the RTFIM, we must finally determine
the value of the exponent $\phi$ which controls the average moment, $\mu \sim |\ln\omega|^\phi$, of
a cluster fluctuating with frequency $\omega$.  This can be accomplished by investigating the
imaginary part of the disorder averaged dynamical order parameter susceptibilities after they have
been analytically continued to real frequencies.  We are interested in the average ($k=0$) and local
susceptibilities defined by
\begin{align}
\label{eq:aveChi}
\mathrm{Im}\; \overline{\chi}(\omega) &= \mathrm{Im} \; \frac{1}{L} \sum_{j} \left. 
\overline{\langle
\Psi^*_j(i\omega)\Psi_0(i\omega)\rangle}_{\mathcal{S}_0}\right|_{i\omega \to \omega + i\epsilon} \\
\mathrm{Im}\; \overline{\chi_{\mathrm{loc}}} (\omega) &= \left. \mathrm{Im}\; \overline{\langle
\Psi^*_j(i\omega)\Psi_j(i\omega)\rangle}_{\mathcal{S}_0}\right|_{i\omega \to \omega + i\epsilon}
\end{align}
where $\langle \cdots \rangle_{\mathcal{S}_0}$ indicates an average with respect to the large-$N$
action in Eq.~(\ref{eq:S0}) as well as a site average.  Note that $\omega$ is now a real
frequency, and we point out that our facile access to such dynamical quantities is one of the 
perquisites of the numerical approach we have taken.  All frequencies are measured
with respect to an ultra-violet cutoff $\Lambda_\omega$ which is required for convergence when
computing the set of solutions to Eq.~(\ref{eq:rj}).
Physically, one can argue that at criticality, the average cluster moment will be given
by the ratio of the average to local susceptibility due to the extra sum over sites
in Eq.~(\ref{eq:aveChi}).
We thus define
\begin{equation}
R(\omega) = \frac{\mathrm{Im} \overline{\chi}(\omega)}{\mathrm{Im} \overline{\chi_{\mathrm{loc}}}
(\omega)} \ \sim \ |\ln \omega|^\phi \ {\cal F}\left( \delta^{\nu \psi} |\ln \omega| \right) \ ,
\label{eq:R}
\end{equation}
and expect that the scaling function ${\cal F}$ approaches a constant when the dimensionless
variable  $\delta^{\nu \psi} |\ln \omega|  \ll 1$. In the quantum disordered phase
with $\delta^{\nu \psi} |\ln \omega|  \gg 1$, a scaling analysis predicts
${\cal F}(x)  \sim  x^{1-\phi}$ \cite{vojta-pc}  and hence
$R  \sim \delta^{\nu\psi(1-\phi)} |\ln\omega|$ ~\cite{fisher,igloi}.
In order to determine the value of $\phi$, it is useful to consider a rescaled value of the
susceptibility ratio $\widetilde{R}(\delta) = R(\omega) / (\delta^{\nu\psi}|\ln\omega|)$ which 
should be frequency independent according to the predicted scaling form for $R(\omega)$ as 
$\omega \to 0$.  We plot the finite size scaled susceptibility ratio in Fig.~\ref{fig:R} for the 
three smallest values of $\delta$, and find confirmation of its $|\ln \omega|$ dependence.
The inset of Fig.~\ref{fig:R} confirms the frequency independence of $\widetilde{R}$
and by determining the best linear fit of $\ln \widetilde{R}$ to $\ln\delta$ for $\omega \le
10^{-3}$ with $\nu \psi=1.0(1)$, we find a cluster exponent $\phi = 1.6(2)$ which is very close
to the predicted RTFIM value of $(1+\sqrt{5})/2$.

The results of the above analysis, as highlighted in Figs.~\ref{fig:eqTimeCorFunc}--\ref{fig:R},
provide compelling evidence for the applicability of the real space RG analysis of Hoyos {\em et 
al.\/}, and further reproduces a number of results of \cite{fisher} to unexpected accuracy. This 
confirms that the considered model for overdamped repulsive Cooperon
fluctuations in the presence of quenched disorder near a SMT exhibits dynamically activated scaling 
and is controlled by an infinite randomness fixed point in the same universality class as the RTFIM.
The transition is characterized by the numerically computed critical exponents
$(\nu,\psi,\phi) \simeq (1.9,0.53,1.6)$ which are entirely consistent with those of
the one dimensional random quantum Ising model in a transverse field $(2,1/2,(1+\sqrt{5})/2)$.

In closing, we note that while our discussion has been framed in the context of the SMT,
models similar to $\mathcal{S}$ describe the onset of a wide variety of orders
in metallic systems \cite{vojta-review}. Furthermore, the flow to the strong-disorder
RTFIM fixed point is expected to also hold in higher dimensions \cite{hoyos}. We thus propose that
our results provide strong support for the applicability of the RTFIM physics to
many experiments involving the onset of spin- and charge-density wave orders in metals.

We thank J.~Hoyos and T.~Vojta for useful discussions.
This research was supported by NSF grants DMR-0537077 and DMR-0605813, the Heisenberg program of 
DFG (BR), and grant PA002-113151 of the SNF (MM).  Computing resources were provided by the
Harvard Center for Nanoscale Systems, part of the National Nanotechnology Infrastructure Network.


\begin{thebibliography}{99}

\bibitem{liu-zadorozhny} Y.~Liu, Yu.~Zadorozhny, M.~M.~Rosario,  B.~Y.~Rock,
P.~T.~Carrigan and
H.~Wang, Science \textbf{294}, 2332 (2001).

\bibitem{lau-markovic} C.~N.~Lau, N.~Markovic, M.~Bockrath,  A.~Bezryadin and
M.~Tinkham Phys. Rev. Lett. \textbf{87}, 217003 (2001).


\bibitem{boogaard} G.~R.~Boogaard, A.~H.~Verbruggen, W.~Belzig, and
T.~M.~Klapwijk,
Phys. Rev. B \textbf{69}, 220503 (2004).

\bibitem{rogachev-bollinger} A.~Rogachev, A.~T.~Bollinger and  A.~Bezryadin, Phys. Rev.
Lett.
\textbf{94}, 017004 (2005).

\bibitem{chang} F.~Altomare, A.~M.~Chang, M.~R.~Melloch, Y.~Hong, and C.~W.~Tu,
Phys. Rev. Lett. \textbf{97}, 017001 (2006).


\bibitem{rogachev-wei} A.~Rogachev, T.-C.~Wei, D.~Pecker,  A.~T.~Bollinger,
P.~M.~Goldart and A.~Bezryadin, Phys. Rev. Lett. \textbf{97}, 137001 (2006).

\bibitem{swt} S.~Sachdev, P.~Werner and M.~Troyer, Phys.  Rev. Lett. \textbf
{92}, 237003
(2004).

\bibitem{mrss} A.~Del Maestro, B.~Rosenow, N.~Shah, and S.~Sachdev, arXiv:0708.0687.

\bibitem{herbut} I.~F.~Herbut, Phys. Rev. Lett. \textbf{85}, 1532  (2000).

\bibitem{lsv} A.~V.~Lopatin, N.~Shah and V.~M.~Vinokur,  Phys. Rev. Lett.
\textbf{94}, 037003
(2005); N.~Shah and A.~V.~Lopatin, arXiv:0705.1890.

\bibitem{spivak-zyuzin} B.~Spivak, A.~Zyuzin and M.~Hruska, Phys.  Rev. B \textbf{64},
132502 (2001).

\bibitem{fl} M.~V.~Feigel'man, A.~I.~Larkin, and M.~A.~Skvortsov,
Phys. Rev. Lett. {\bf 86}, 1869 (2001); V.~M.~Galitski
and A.~I.~Larkin, Phys. Rev. Lett. {\bf 87}, 087001 (2001).

\bibitem{galitski} V.~Galitski, arXiv:0708.3841; arXiv:0710.1868.

\bibitem{vojta-review} T.~Vojta, J. Phys. A \textbf{39}, R143 (2006).

\bibitem{hoyos} J.~A.~Hoyos, C.~Kotabage, and T.~Vojta, Phys. Rev. Lett. {\bf 99}, 230601 (2007).

\bibitem{fisher} D.~S.~Fisher, Phys. Rev. Lett. {\bf 69}, 534 (1992); Phys. Rev. B {\bf 51}, 6411
(1995).

\bibitem{tucker} J.~R.~Tucker and B.~I.~Halperin, Phys. Rev. B \textbf{3}, 3768 (1971).

\bibitem{satya} T.~Senthil and S.~N.~Majumdar,
Phys. Rev. Lett. {\bf 76}, 3001 (1996).

\bibitem{tu-weichman} Y.~Tu and P.~B.~ Weichman, Phys. Rev. Lett \textbf{73}, 6 (1994).

\bibitem{hartman-weichman} J.~W.~Hartman and P.~B.~Weichman, Phys. Rev. Lett \textbf{74}, 4584 
(1994).

\bibitem{griffiths} R.~B.~Griffiths, Phys. Rev. Lett. \textbf{23}, 17 (1969).


\bibitem{young-rieger} A.~P.~Young and H.~Rieger, Phys. Rev. B \textbf{53}, 8486 (1996).

\bibitem{vojta-pc} T.~Vojta, private communication.

\bibitem{igloi} F.~Igl\'{o}i, Phys. Rev. B \textbf{65},064416 (2002); F.~Igl\'{o}i and C.~Monthus,
Phys. Rep. \textbf{412}, 277 (2005).


\end{thebibliography}
\end{document}